\documentclass[pdflatex,sn-mathphys-num]{sn-jnl}

\usepackage{graphicx}%
\usepackage{multirow}%
\usepackage{amsmath,amssymb,amsfonts}%
\usepackage{amsthm}%
\usepackage{mathrsfs}%
\usepackage[title]{appendix}%
\usepackage{xcolor}%
\usepackage{textcomp}%
\usepackage{manyfoot}%
\usepackage{booktabs}%
\usepackage{algorithm}%
\usepackage{algorithmicx}%
\usepackage{algpseudocode}%
\usepackage{listings}%
\usepackage{lineno}
\usepackage{hyperref}
\usepackage{graphicx}
\usepackage[export]{adjustbox}

\theoremstyle{thmstyleone}%
\theoremstyle{thmstyletwo}%

\theoremstyle{thmstylethree}%

\begin{document}

\title[Article Title]{Confinement inhibits surficial attachment and induces collective behaviors in bacterial colonies}


\author*[1,2,3]{\fnm{Vincent} \sur{Hickl}}\email{vincent.hickl@empa.ch}

\author[1,2,3]{\fnm{Gabriel} \sur{Gm\"under}}

\author[2]{\fnm{Ren\'e M.} \sur{Rossi}}

\author[3]{\fnm{Antonia} \sur{Neels}}

\author[1]{\fnm{Qun} \sur{Ren}}

\author[1]{\fnm{Katharina} \sur{Maniura-Weber}}

\author[1,2,3]{\fnm{Bruno F. B.} \sur{Silva}}

\affil*[1]{\orgdiv{Laboratory for Biointerfaces}, \orgname{Empa}, \orgaddress{\street{Lerchenfeldstrasse 5}, \city{St. Gallen}, \postcode{9000}, \country{Switzerland}}}

\affil[2]{\orgdiv{Laboratory for Biomimetic Membranes and Textiles}, \orgname{Empa}, \orgaddress{\street{Lerchenfeldstrasse 5}, \city{St. Gallen}, \postcode{9000}, \country{Switzerland}}}

\affil[3]{\orgdiv{Center for X-ray Analytics}, \orgname{Empa}, \orgaddress{\street{Lerchenfeldstrasse 5}, \city{St. Gallen}, \postcode{9000}, \country{Switzerland}}}


\abstract{Bacterial colonies are a well-known example of living active matter, exhibiting collective behaviors such as nematic alignment and collective motion that play an important role in the spread of microbial infections. While the underlying mechanics of these behaviors have been described in model systems, many open questions remain about how microbial self-organization adapts to the variety of different environments bacteria encounter in natural and clinical settings. Here, using novel imaging and computational analysis techniques, the effects of confinement to 2D on the collective behaviors of pathogenic bacteria are described. Biofilm-forming \textit{Pseudomonas aeruginosa} are grown on different substrates, either open to the surrounding fluid or confined to a single monolayer between two surfaces. Orientational ordering in the colony, cell morphologies, and trajectories are measured using single-cell segmentation and tracking. Surprisingly, confinement inhibits permanent attachment and induces twitching motility, giving rise to multiple coexisting collective behaviors. This effect is shown to be independent of the confining material and the presence of liquid medium. The nematic alignment and degree of correlation in the cells' trajectories determines how effectively bacteria can invade the space between two surfaces and the 3D structure of the colony after several days. Confinement causes the formation of dynamic cell layers driven by collective motion as well as collective verticalization leading to the formation of densely packed crystalline structures exhibiting long-range order.  These results demonstrate the remarkable breadth of collective behaviors exhibited by bacteria in different environments, which must be considered to better understand bacterial colonization of surfaces.}


\maketitle

\section*{Main}\label{sec1}

\subsection*{Introduction}

Bacterial colonies are a well-known example of active matter, where individual cells convert chemical energy into mechanical work, leading to a plethora of non-equilibrium physical phenomena~\cite{DellArciprete2018,Doostmohammadi2018,Aranson2022}. Their self-organization has been studied extensively for its importance in biological processes like infectious diseases~\cite{Costerton1999}, agriculture~\cite{Rudrappa2008}, biofouling~\cite{Flemming2020}, and bioremediation of environmental pollutants~\cite{Singh2006,Hazen2015,Hickl2023}. The growth of bacteria at surfaces is especially important, and is typically described in terms of several distinct stages, including the planktonic state, irreversible surface adhesion, microcolony formation, biofilm maturation, and dispersion~\cite{Flemming2010,Rather2021,Sauer2022}. Throughout all these stages, collective behaviors of bacterial cells play a crucial role in determining how the colony develops. In particular, orientational ordering and collective motility of cells determine how bacteria spread across surfaces, and how two-dimensional microcolonies eventually become three-dimensional biofilms~\cite{Yan2016, Hartmann2019, Shimaya2022}. Thus, quantitative descriptions of these phenomena is fundamental to understanding the biophysics of bacterial proliferation.

A variety of collective behaviors have been described in colonies of different bacterial species growing at surfaces with different geometries, material properties, and nutrient sources. Monolayers of growing non-motile \textit{Escherichia coli} undergo an isotropic-to-nematic transition in time as they become densely packed in microfluidic channels~\cite{Volfson2008}. In \textit{Vibrio cholerae} biofilms and \textit{P. aeruginosa} colonies, orientational ordering and steric forces lead to cell verticalization and distinctive 3D structure in mature biofilms~\cite{Beroz2018, Meacock2020}. In \textit{Myxococcus xanthus} colonies, collective motion driven by topological defects causes layer formation, which can be explained entirely by cell motility and mechanical interactions~\cite{Copenhagen2020,Han2025}. Both \textit{M. xanthus} and \textit{B. subtilis} also undergo a phase transition resembling motility-induced phase separation to form dense clusters and fruiting bodies~\cite{Liu2019,Grobas2021}. These different behaviors have been described physically by individual theoretical and computational models~\cite{You2018,Nagel2020,You2021,Langeslay2023a}, but a general description of bacterial collective behaviors remains elusive because of the many physical, chemical, and biological factors that influence their self-organization~\cite{Bechinger2016,Arnaouteli2021,Sauer2022,Moore-Ott2022}. In order to predict how bacteria with different properties will behave in different environments, new experiments are needed to decouple these factors and describe how each of them influences bacterial collective behaviors.

Recent work has revealed that different physical factors, including material properties of the substrate and the geometry of the bacteria's environment affect how bacteria grow and proliferate at surfaces~\cite{Cho2007}. Confinement is particularly important, as bacteria must contend with restricted spaces in many natural systems, including porous materials like soil, oceanic sediments, and subsurface environments that host a large majority of the earth's microorganisms. Human pathogens similarly invade confined spaces, including capillary blood vessels, mucosal interfaces, and interfaces between implants and tissues. Confinement in rectangular channels has been shown to induce long-range orientational ordering of cells in 2D~\cite{Volfson2008,Cho2007}. Individual bacterial swimming patterns and speeds have been shown to change in response to confinement in porous media~\cite{Bhattacharjee2019} and between two flat surfaces~\cite{Chen2024}. The latter mode of confinement, where cells are physically confined to a quasi-infinite 2D plane, is often used to facilitate high-resolution imaging, but, despite its importance, its effect on the collective behaviors of bacteria has not been characterized systematically. 

Here, we describe novel results on collective behaviors induced confinement of bacteria to two dimensions. Bacteria are grown confined and unconfined at solid-liquid, solid-air, and solid-solid surfaces to decouple the effect of confinement from the type of surface. High resolution microscopy and quantitative image analysis, including single-cell segmentation and tracking, are used to quantify the physical properties of the resulting active nematic, including orientational ordering, collective motion, and single-cell morphologies. Confinement is shown to inhibit permanent surface attachment, induce twitching motility, and give rise to two distinct modes of collective motion that can coexist under a single set of experimental conditions. The degree of orientational ordering and collective motion in these modes determines the architecture of the bacterial colony in 3D. Groups of cells unable to effectively invade the confined space become verticalized, forming dense, highly ordered crystalline structures, while those able to sustain colony expansion form multiple dynamic cell layers. These results provide important insights into the role of system geometry on biological active matter, and have important implications for our understanding of the study of infections in physiologically relevant environments.

\subsubsection*{Self-organization of unconfined \textit{P. aeruginosa} at surfaces}\label{subsec1}

Unconfined growth of \textit{Pseudomonas aeruginosa} (\textit{P. a.}) at surfaces leads to the formation of microcolonies of permanently attached, non-motile cells. Colonies of \textit{P. a.} were grown in a microfluidic flow chamber in tryptic soy broth (TSB) and imaged using timelapse confocal microscopy (Fig 1A). Over time, bacteria transition from the swimming, planktonic state in the bulk liquid to the surface-attached, sessile state. Bacterial attachment to the substrate is initially reversible, as individual cells sometimes leave the surface after initially being attached (supporting video 1). As  cells attach permanently to the solid-liquid (glass-TSB) interface and proliferate, their surficial motility is negligible compared to their planktonic counterparts. The bacteria thus form 2D microcolonies of statically attached cells at the surface which eventually become 3D biofilms (Fig 1B). Similarly, when grown at an air-solid interface on nutrient-rich tryptic soy agar (TSA) (Fig 1C), bacteria are non-motile and grow in dense, permanently attached colonies (Fig 1D). In this case, any motion of bacteria across the surface is very slow, and is caused by steric interactions between growing cells, rather than motility (supporting video 2). In principle, \textit{P. a.} are capable of both flagella- and pili-mediated motility in bulk liquid and at surfaces, respectively. In this case, while flagellar motility is observed consistently in liquid growth media, attachment at unconfined surfaces did not lead to pili-mediated ``twitching" motility. This result is surprising, as twitching motility of surface-attached \textit{Pseudomonas} is routinely assumed in the literature, and has been described as a prerequisite of biofilm formation~\cite{OToole1998}. Here, it is shown that a \textit{wild type} (\textit{WT}) strain does not exhibit twitching motility at surfaces under two very different sets of conditions, suggesting greater care is needed when describing how bacteria respond to surface-attachment. In both the setups described here, microcolony growth eventually lead to biofilm formation.

\begin{figure*}
  \centering
  \includegraphics[width=\linewidth]{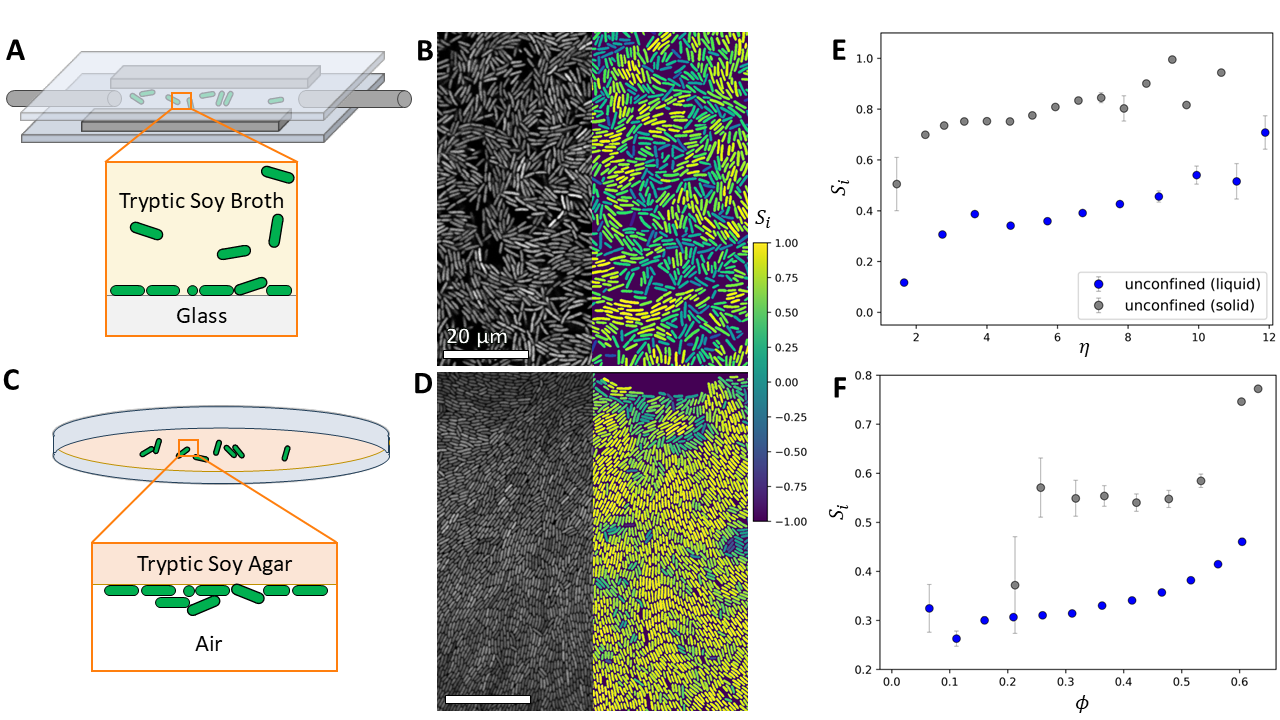}
  \caption{{Unconfined monolayers.} (A) Diagram of microfluidic flow chamber for bacterial growth on unconfined surfaces in liquid growth medium tryptic soy broth (TSB). (B) Representative microcolony of unconfined \textit{P. aeruginosa} on glass in liquid growth medium. Cells on the right side are colored by their nematic order parameter $S_i$. (C) Diagram of unconfined bacteria colony growing at a solid-air interface on solid growth medium tryptic soy agar (TSA) (D) Representative microcolony of unconfined \textit{P. a.} on solid TSA, cells on the right are colored by $S_i$. (E) Single-cell nematic order parameter $S_i$ as a function of cell aspect ratio $\eta$ for unconfined \textit{P. a.} colonies. (F) Nematic order parameter as a function of cell packing fraction $\phi$ for unconfined \textit{P. a.}.}
  \label{fig1:unconfined}
\end{figure*}

The bacteria's orientational ordering in unconfined, surface-attached microcolonies depends on the type of interface they are growing on. Using single-cell segmentation, the orientation of each cell in a monolayer (prior to biofilm formation) can be determined (Fig. 1B). For each pair of nearby cells (cells within $20$ \textmu m of one another), the alignment is quantified as
\begin{equation}
    S_{ij} = 2\cos^2{(\theta_i-\theta_j)-1}
\end{equation}
where $\theta_i$ and $\theta_j$ are the orientations of the two cells. This quantity is  equal to $1$ for parallel alignment and $-1$ for perpendicular alignment (note that the maximum angle difference is $\pi/2$, as the rod-shaped cells have head-to-tail symmetry). Nearby cells tend to align, and the alignment decays exponentially with increasing cell-to-cell distance $r_{ij}$ (Fig~\ref{figS1:unconfined}C). The decay constants are $1.81\pm0.04$ \textmu m and $25.4\pm0.7$ for microcolonies on solid-liquid and solid-air interfaces, respectively, meaning cell orientations are correlated for distances an order of magnitude larger on solid agar. This decay constant is equivalent to the mean correlation length of cell orientations. The average of $S_{ij}$ over all of each cell's neighbors is calculated, giving the single-cell nematic order parameter $S_i$, which is a measure of each cell's alignment:
\begin{equation}
    S_i = \frac{1}{N}\sum_{j\in C_r}2\cos^2{(\theta_i-\theta_j)-1}
\end{equation}
The sum is taken over all cells in a circular area of radius $r$ around the $i^{\text{th}}$ cell. $N$ is the number of cells in that area. Here, $r$ is chosen to equal the mean cell length, $\bar{l}=3.2$ \textmu m. Bacteria grown on the solid-air (TSA-air) interface are significantly more ordered than bacteria grown on a solid-liquid (glass-TSB) interface (Fig 1B\&D, right). 

At the single cell scale, orientational order in surface-attached microcolonies depends on the packing fraction and cell aspect ratio. Using single-cell segmentation, the aspect ratio of each cell $\eta = l/w$, where $l$ and $w$ are the cell length and width, and the local packing fraction $\phi$ -- the fraction of the surface covered by bacteria -- are measured. Under both conditions, the nematic order parameter increases with increasing cell aspect ratio (Fig 1E). This observation is consistent with the theory of densely-packed rods, where the nematic phase is favored for particles with high aspect ratios, and is in qualitative agreement with previous theoretical and experimental work~\cite{Weitz2015,Sheats2017}. However, thanks to the state-of-the-art single-cell segmentation method used here, it is the first demonstration of the effect of cell elongation on orientational ordering of individual cells. Under both experimental conditions, $S_i$ does not depend strongly on $\phi$ at low packing fractions ($\phi\lesssim0.5$), (Fig 1F). Then, $S_i$ increases sharply for more densely packed cells, especially for colonies grown on TSA. This upturn suggests the existence of a transition to a more highly ordered nematic state, as is expected from the theory of liquid crystals~\cite{Onsager1949,Flory1956}. However, no truly isotropic state was found, as $S_i>0$ for all packing fractions, which is likely due to the fact that bacteria divide end-to-end, and thus tend to align even when not densely packed. The statistics of bacterial self-organization for unconfined colonies on solid-liquid surfaces is not sensitive to surface properties. In additional experiments, bacteria were grown on polydimethylsiloxane (PDMS) rather than glass, which resulted in nearly identical orientational ordering of the colonies (Fig A1). Note that PDMS is several orders of magnitude softer than glass and significantly more hydrophobic.

\subsection*{Modes of collective behaviors under confinement}

Confinement of bacteria between two surfaces inhibits permanent attachment and induces motility, leading to distinct modes of collective behaviors. To test the effect of confinement on bacterial attachment and growth, \textit{P. a.} colonies were grown from a few cells to dense colonies between PDMS and glass (suspended in a thin film of TSB) or between glass and TSA (without liquid media), as shown in Fig 2A. In both cases, bacteria are confined entirely in 2D and can be imaged continuously as they proliferate and self-organize into densely-packed monolayers. Unlike in the unconfined growth on glass or PDMS described in Fig 1, bacteria do not become non-motile when in contact with the surfaces between which they are confined. Instead, cells exhibit sustained motility between the two surfaces (see supporting videos 1 and 2). This motility occurs in the presence of liquid medium (between PDMS and glass) and in its absence (between glass and TSA), as shown by the trajectories in Fig 2B-C. In some cases, single-cell trajectories appear entirely uncoordinated, qualitatively resembling random walks (Fig 2B). This mode is henceforth referred to as ``chaotic twitching.'' In other cases, trajectories appear highly coordinated, with groups of cells moving collectively in the same direction (Fig 2C), which is henceforth referred to as ``collective twitching.'' Confining \textit{P. a.} between PDMS and glass exclusively induces chaotic twitching, whereas confinement between glass and TSA can induce chaotic twitching or collective twitching. In the latter case, the two distinct behaviors can occur in different microcolonies within a single sample, meaning \textit{P. aeruginosa} can exhibit both modes of collective behavior under identical conditions.

The motility induced by confinement is due to bacterial twitching using type 4 pili (T4P). Additional experiments were carried out with \textit{pilA} mutants, a strain in which twitching motility was disabled. This mutant strain exhibits swimming motility due to flagellar propulsion in liquid medium (supporting video 3). However, when confined between glass and TSA, \textit{pilA} mutants exhibit neither chaotic nor collective twitching (supporting video 4). Instead, colony development closely resembled unconfined WT bacteria growing on agar undergoing non-motile growth, with cells moving only because of steric forces caused by cell division. Previous work has shown that, when \textit{P. a.} encounter surfaces, consequent biomechanical signaling can suppress twitching motility, causing permanent surface attachment and biofilm formation. Surprisingly, confinement between two surfaces seems to reverse this mechanism, inhibiting the suppression of T4P-based twitching motility, and thus giving rise to chaotic and collective twitching.

\begin{figure*}
  \centering
  \includegraphics[width=\linewidth]{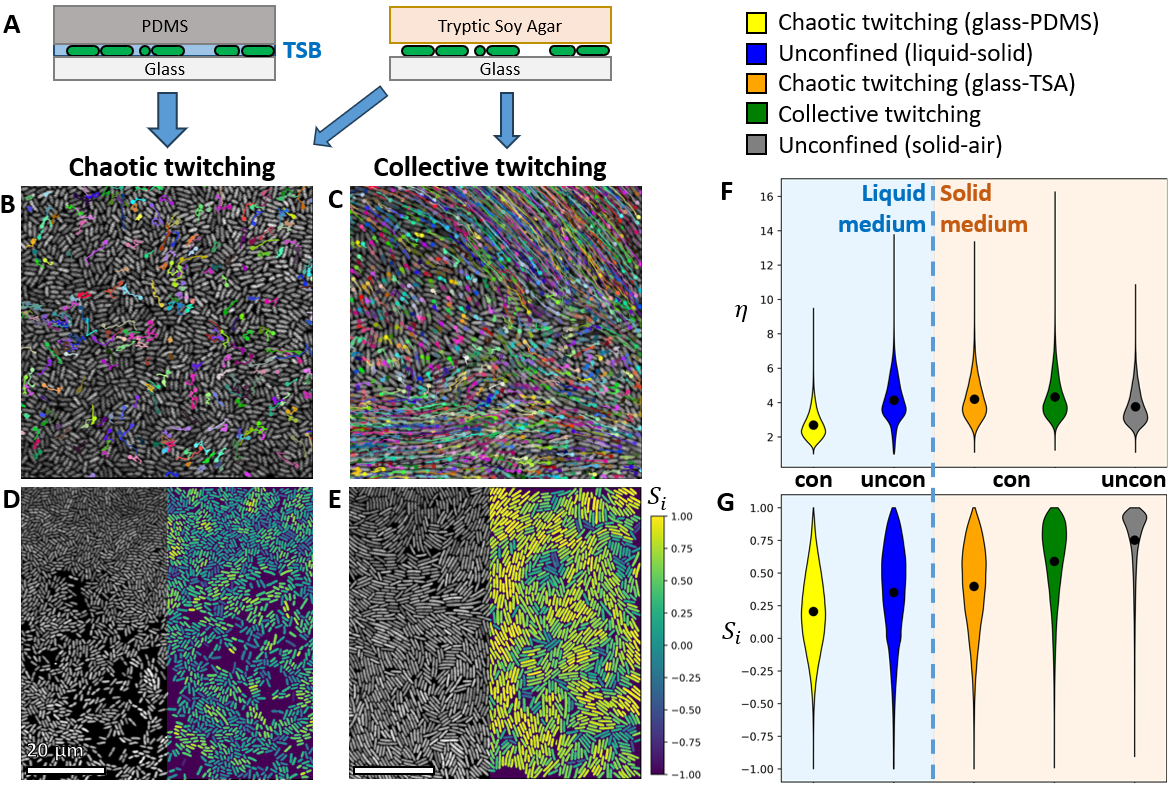}
  \caption{{Modes of collective motion under confinement.} (A) Experimental conditions leading to distinct collective behaviors under 2D confinement: confinement between PDMS and glass or between TSA and glass. (B)-(C) representative single-cell trajectories for chaotic twitching and collective twitching, respectively. Colored dots and lines are subsets of trajectories over 1-3 minutes (see supporting videos 5 and 6). (D) Representative image of bacteria exhibiting chaotic twitching. Right half shows single-cell nematic order parameters. (E) Bacteria exhibiting collective twitching, along with nematic order parameters. (F) Distributions of single-cell aspect ratios from different modes of collective behaviors for unconfined and confined bacteria. Black dots represent mean values. (G) Distributions of single-cell nematic order parameters from different modes of collective behaviors and experimental conditions. Black dots represent mean values. For each distribution in F and G, $N>20000$ cells. Scale bars are 20 \textmu m.}
  \label{fig2:phases}
\end{figure*}

Cell morphologies depend on the type of confinement and the surface composition. In liquid medium, confined cells (between glass and PDMS) have significantly smaller aspect ratios (mean value $\Bar{\eta} = 2.690\pm0.002$) compared to unconfined cells on glass ($\Bar{\eta} = 4.134\pm0.003$). On the other hand, on solid medium, confined cells (between glass and TSA) have significantly higher aspect ratios ($\Bar{\eta} = 4.198\pm0.007$ and $\Bar{\eta} = 4.322\pm0.006$ for chaotic twitching and collective twitching, respectively) compared to unconfined cells on TSA ($\Bar{\eta} = 3.754\pm0.007$), as shown in Fig 2F. 

The observed differences in aspect ratios cannot be explained by nutrient availability, the material properties of the surfaces, or differences between liquid and solid nutrient sources alone. Generally, rod-shaped can change their size and shape to adapt to different nutrient sources and availabilities~\cite{Henrici1928,Steinberger2002,Ojkic2021}. This phenomenon could explain why cells confined between PDMS and glass have much shorter aspect ratios. However, the reduction in aspect ratio was observed while bacteria were still swimming and proliferating, suggesting at least some nutrient availability. Additionally, nutrient depletion cannot explain why unconfined cells undergoing non-motile growth on TSA also had lower aspect ratios, especially compared to confined cells also growing on TSA. Unconfined bacteria are shorter when grown on solid TSA, but confined bacteria are shorter when grown between glass and PDMS in liquid TSB. Thus, our results suggest that cells adapt their shape to their environment in highly complex ways based on several environmental cues. In particular, 2D confinement itself plays an important role in determining cell morphologies. The biophysical mechanism underlying this effect will require further study.

Orientational ordering of bacterial colonies is significantly reduced by 2D confinement. When confined between glass and PDMS in liquid TSB, the nematic order is significantly smaller ($\Bar{S_i} = 0.205$) compared to the unconfined growth on glass or PDMS ($\Bar{S_i} = 0.351$), as shown in Fig 2I. Similarly, between glass and solid TSA, alignment is significantly reduced ($\Bar{S_i} = 0.49$) compared to unconfined growth on TSA ($\Bar{S_i} = 0.751$). Additionally, cells undergoing collective twitching are more aligned ($\Bar{S_i} = 0.589$) compared to those undergoing chaotic twitching ($\Bar{S_i} = 0.397$). 
The motility induced by confinement means the bacteria move on much shorter time scales, increasing the activity in the monolayer and thus contributing to the reduction in nematic ordering. However, \textit{P. a.} can greatly increase their orientational order by entering the collective twitching mode.

\subsection*{Quantifying modes of bacterial self-organization}

\begin{figure*}
  \includegraphics[width=\linewidth, left]{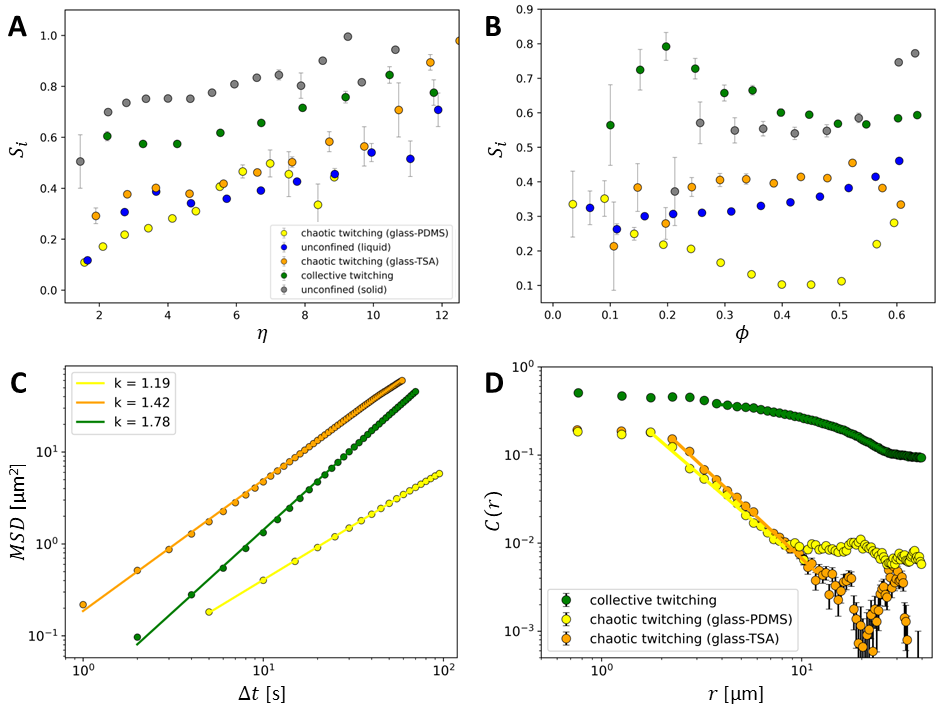}
  \caption{{Quantitative description of collective behaviors.} (A) Single-cell nematic order parameter vs. cell aspect ratio for different modes and conditions. Error bars represent standard errors, legend applies to A and B. (B) Single-cell nematic order parameter vs. cell packing fraction. (C) Mean square displacements as a function of the lag. Lines represent best power law fits, with exponents given in the legend. (D) Spatial velocity correlation functions $C(r)$ vs. cell-cell distance $r$. Lines represent best power law (chaotic twitching) or exponential (collective twitching) fits. Legend applies to both C and D.}
  \label{fig3:statistics}
\end{figure*}

The reduction in orientational ordering of confined bacteria is driven by the increase in the bacteria's motility and, in liquid medium, by the reduction in cell aspect ratio. Under all conditions, the nematic order parameter $S_i$ increases with increasing cell aspect ratio (Fig 3A). When comparing unconfined growth in liquid medium with chaotic twitching under confinement, the difference in alignment can be explained mostly by differences in the aspect ratio. At any particular aspect ratio, there is little difference in the mean value of $S_i$ (as shown by the orange, yellow, and blue curves). Thus, the lower alignment under confinement between PDMS and glass is best explained by the lower aspect ratios of cells under those conditions. Notably, the dependence of $S_i$ on $\eta$ is approximately the same for chaotic twitching between glass and PDMS (with liquid TSB) and between glass and TSA, suggesting the motion induced by confinement is the dominant factor, rather than any specific interactions with the surfaces or the growth media. On the other hand, the much higher alignment observed for collective twitching cannot be explained by differences in the cell aspect ratio. Rather, it is the result of the distinctive collective motion that arises in this state, leading to an increase in orientation order compared to chaotic twitching. At all aspect ratios $\eta>2$, the confinement-induced motility causes cells in undergoing collective twitching to be less ordered than cells grown unconfined on TSA. Overall, nematic alignment in bacterial colonies depends on the complex interplay of confinement, growth conditions, cell morphologies, and cell motility.

No transition to a more ordered nematic phase is observed under confinement between glass and TSA. Unlike for unconfined colonies, $S_i$ does not increase with the packing fraction $\phi$ at higher densities. For collective twitching, $S_i$ is highest for relatively low packing fractions ($\phi\approx0.2$), and decreases thereafter until it reaches a plateau at $\phi\gtrsim0.5$. Bacteria actively align even when not densely packed, which could contribute to their collective motion, and represents a significant deviation from the behavior of active rods with only steric interactions~\cite{Kayser1978}. This result suggests there may be a biochemical interaction induced by confinement which favors alignment and collective motion. For chaotic twitching between glass and TSA, there is no strong dependence of $S_i$ on $\phi$, which also represents a deviation from theoretical expectations, but in a different manner than collective twitching. For confined bacteria between glass and PDMS undergoing chaotic twitching, a steady decrease in ordering with increasing packing fraction is observed for $\phi\lesssim0.4$, followed by a sharp increase at higher densities. This phenomenon represents yet another unique behavior, but includes the expected transition to a higher ordered state at high densities. Together, these results show that the active matter physics of bacterial-self-organization under confinement are not explained by simple models of rigid active rods reported in the literature. Significant further theoretical and experimental work is needed to better capture the dynamics of bacterial alignment under diverse conditions to explain their dependence on 2D confinement and cell densities.

Modes of collective motion can be characterized by different mean square displacements and spatial velocity correlation functions. Mean square displacement ($MSD =  \langle |\Vec{x}(\Delta t) - \Vec{x}(0)|^2 \rangle$) is computed from single-cell tracks of bacteria in fast timelapses (0.2 - 1 fps) of their motion obtained using confocal microscopy. Average values of $MSD$ from many single-cell tracks are plotted as a function of lag time $\Delta t$, showing a power law dependence (Fig 3C):
\begin{equation}
    \langle |\Vec{x}(\Delta t) - \Vec{x}(0)|^2 \rangle = \alpha \Delta t^\beta
\end{equation}
The power law exponent $\beta$ can be used to characterize the trajectories, with $\beta = 1$ corresponding to diffusive (Brownian) motion, and $\beta = 2$ corresponding to straight-line (ballistic) motion. Confined bacteria undergoing chaotic twitching have lower exponents, ranging from $1.19$ (between PDMS and glass with liquid medium) to $1.42$ (between glass and solid TSA), demonstrating the relatively diffusive motion in this mode, and suggesting that the presence of liquid medium or the material properties of the confining surfaces play a role in determining how bacteria move under confinement. Confined bacteria undergoing collective twitching (between glass and solid TSA) exhibit motion that is significantly less random and more directed, with an exponent of $1.78$. Thus, the increased ordering under collective twitching allows bacteria to have more directed motion, which could contribute to their ability to explore the interface at which they are confined.

The spatial velocity correlation function $C(r)$ is calculated based on the instantaneous velocities of the bacteria $v_i$:
\begin{equation}
    C(r_{ij}) = \frac{\langle \Vec{v_i}\cdot\Vec{v_j} \rangle}{\langle v^2\rangle}
\end{equation}
where $r_{ij} = |\vec{r_i} - \vec{r_j}|$ is the distance between cells, and $\langle v^2 \rangle$ is the mean velocity squared over all tracked cells. The average in the numerator is taken over all cell pairs separated by each distance $r_{ij}$ (after binning the data). Under chaotic twitching, regardless of the confining materials and whether liquid medium is present, neighboring cells ($r_{ij} < 2$ \textmu m) have weakly correlated velocities ($C(r)\approx 0.2$), and the correlation function decays quickly with increasing distance up to $r_{ij}\approx10$ \textmu m. In this intermediate regime, the decay follows a power law relation similar to Eq. 3:
\begin{equation}
    C(r_{ij}) = \gamma {r_{ij}}^{\delta}
\end{equation}
with best fit parameters $\gamma = 0.47\pm0.02$ and $\delta = -1.80\pm0.07$. Notably, the correlation function does not decay below $0$, even for distances up to $40$ \textmu m. By contrast, under collective twitching, neighboring cells have strongly correlated velocities ($C(r)\approx 0.5$). Additionally, the decay of the correlation function is exponential rather than following a power law:
\begin{equation}
    C(r_{ij}) = \gamma' e^{\delta' r_{ij}}
\end{equation}
with $\gamma' = 0.46$ and $\delta' = 0.56$, which corresponds to a much lower spatial rate of decay compared to chaotic twitching. 
Thus, the different modes of collective behaviors can be rigorously differentiated not only by the magnitude of the alignment and the cell swimming speeds, but also by the power law exponent of the mean square displacement, and by the functional form of the decay of the spatial velocity correlation function.

\subsection*{Mass cell verticalization caused by chaotic twitching}

\begin{figure*}
  \includegraphics[width=\linewidth, left]{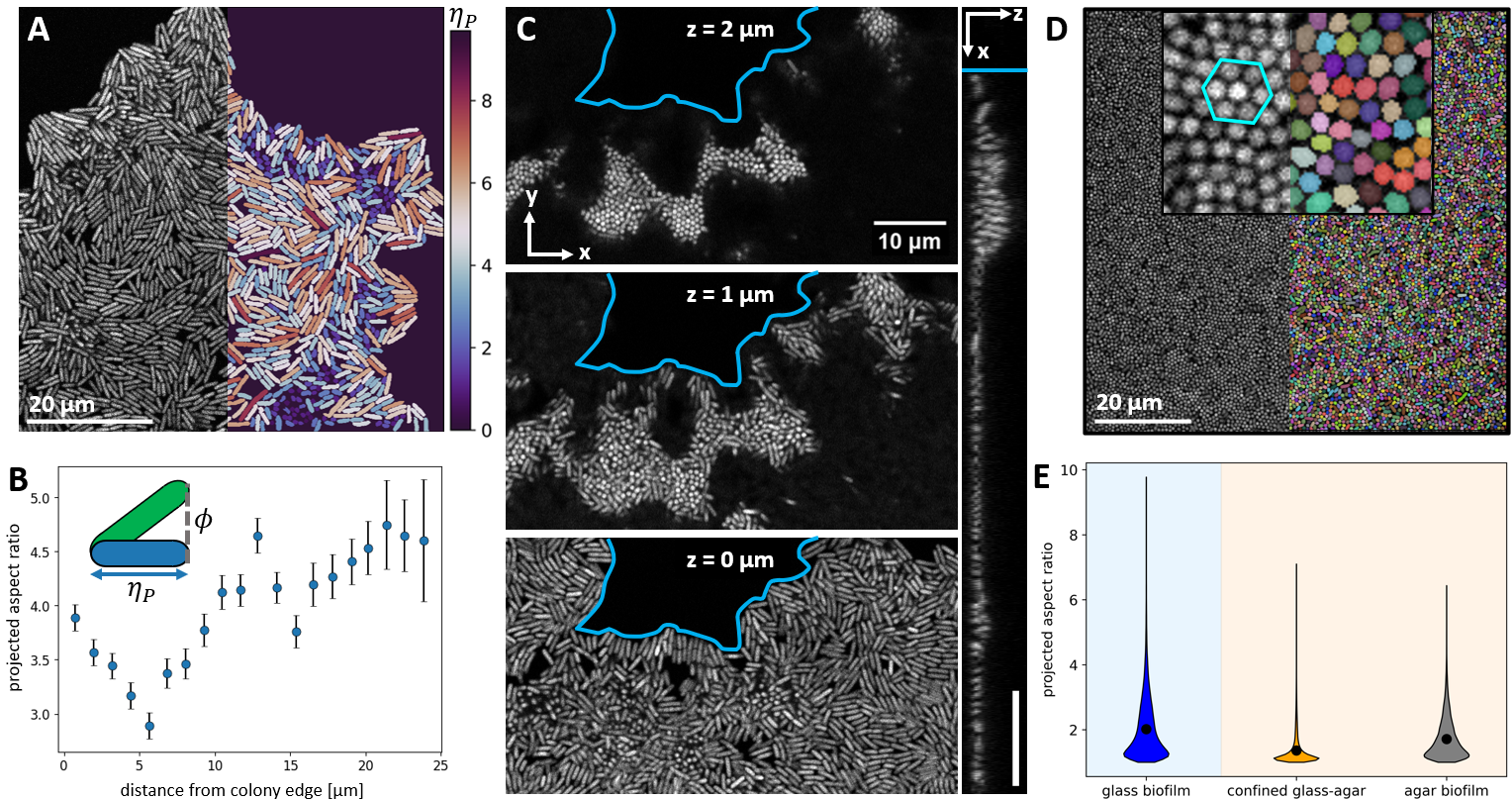}
  \caption{{Chaotic twitching leads to mass verticalization.} (A) Edge of colony undergoing chaotic twitching. Right side: segmented cells colored by their projected aspect ratio (the aspect ratio of each cell in the focal plane). (B) Projected aspect ratios $\eta_P$ as a function of the distance from the edge of the colony, measured only from colonies undergoing chaotic twitching. (C) Snapshots from z-stack of colony undergoing chaotic twitching after about 20 hrs, with focal planes $1$ \textmu m apart. Verticalized cells are shown breaking out of the monolayer. Right: representative orthogonal slice from a z-stack showing verticalization near the edge of the colony. (D) Representative image of colony after 36-48 hrs, with nearly all cells having tilted vertically upwards. Right side shows single-cell segmentation masks (random colors). Inset: closeup of densely-packed vertical cells exhibiting highly ordered, hexagonal ordering. (E) Probability density function of projected aspect ratio $\eta_P$ in unconfined biofilms grown on glass (in liquid medium) or TSA (solid medium) and confined colonies between glass and TSA. Confined colonies exhibit mass verticalization resulting in smaller values of $\eta_P$. See Fig S2 for representative images.}
  \label{fig4:verticalization}
\end{figure*}

After confined bacteria form densely-packed monolayers (about 18-24 hrs after the start of the experiments presented here), the further development of their colonies differs significantly based on the type of confinement and the mode of collective behavior. Between PDMS and glass, nutrient depletion of the liquid medium causes bacteria to stop moving and proliferating within a few hours of when a dense monolayer forms. On the other hand, between glass and TSA, bacteria continue to proliferate thanks to the abundant nutrients in the agar, and colonies expand across the glass-TSA interface. Under chaotic twitching, bacteria in the interior of each colony are significantly more mobile than cells near the edge (supporting video 6). Within 18-24 hours, cells directly behind the leading edge of the colony begin tilting up from the interface (Fig 4A). Their tilt can be measured indirectly via the projected aspect ratio $\eta_P$, which is the apparent length of the cell in the focal plane divided by its width, and is significantly lower for cells $2-10$ \textmu m from the edge of the colony (Fig 4B). This verticalization of cells near the colony edge can be clearly observed in z-stacks of the colony (Fig 4C), which also show that the rest of the cells continue to form a 2D monolayer at this time. Over the subsequent 12-24 hours, nearly all cells throughout the colony become verticalized, forming a dense and highly ordered crystalline lattice (Fig 4D). Vertically oriented cells are hexagonally packed (Fig 4D, inset), which is the densest possible packing of such rods. Thus, chaotic twitching over extended time periods leads to cell verticalization in confined bacterial colonies. After cells become verticalized and densely packed, they stop exhibiting twitching motility at the interface. 

The resulting structure of the now 3D confined colony is significantly different from unconfined biofilms grown on both glass in TSB and on TSA. In all three systems, $\eta_P$ is measured for all cells in representative slices of the colony $\sim2$ \textmu m above the surface (Fig 4E). In unconfined biofilms, cell exhibit a broad range of orientations, whereas under confinement the vast majority of cells becomes verticalized (see Fig A2). Assuming the distribution of cell aspect ratios remains the same that was measured in cell monolayers (Fig 2F), the mean orientations of cells with respect to the surface, $\phi$, can be estimated. The mean angle between the cell and the surface is $60.7^\circ$, $62.7^\circ$, and $71.3^\circ$ respectively for biofilms grown unconfined on glass and agar and confined between the two. 

\subsection*{Layer formation caused by collective twitching}

\begin{figure*}
  \includegraphics[width=\linewidth, left]{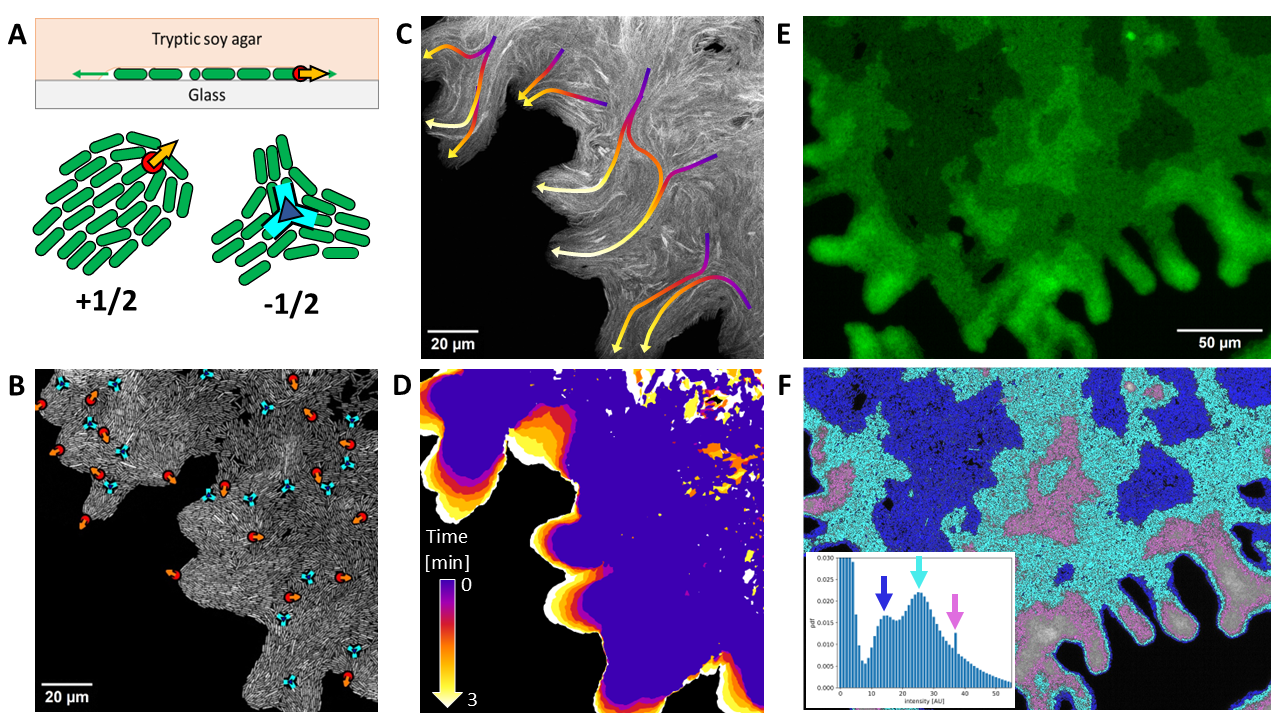}
  \caption{{Collective twitching leads to layer formation.} (A) Top: schematic of confined cells pushing against the glass-TSA interface during colony expansion. Bottom: diagram of $+1/2$ and $-1/2$ topological defects. (B) Representative confocal image of colony edge undergoing collective twitching, with $\pm1/2$ topological defects shown in red and blue, respectively. (C) Maximum intensity projection over time obtained from a video of colony expansion under collective twitching. Arrows represent major collective flows in the colony (see supporting video 7). (D) Outlines of the colony at successive time points $30$ s apart showing outward colony expansion. (E) Widefield fluorescence image of colony during layer formation. Brighter areas correspond to more cell layers. (F) Widefield image colored by the number of cell layers. Inset: histogram of pixel intensities, showing 3 distinct peaks corresponding to different numbers of layers.}
  \label{fig5:layers}
\end{figure*}    

Under collective twitching, confined bacterial monolayers contain persistent topological defects, particularly at the edges of the colony. Topological defects are points where the orientational ordering of the bacteria breaks down, and are characterized by the winding number of the director field about that point. The two most common defect types in bacterial colonies are $\pm1/2$ defects, corresponding to ``comet'' and ``aster" shapes, respectively (Fig 5A). Such defects are present throughout cell monolayers, but, under collective twitching, $+1/2$ defects notably appear at the ends of groups of bacteria which protrude from the colony, as shown in Fig 5B. Taking the maximum intensity projection over time of a timelapse of bacterial motility (supporting video 7) allows the trajectories of bacteria through the colony to be visualized (Fig 5C), and shows that bacteria move collectively towards the protrusions at the edge. These protrusions then drive the expansion of the colony along the glass-TSA interface (Fig 5D). This result suggests that collective flows and topological defects play an important role in the spreading of confined bacteria along an interface.

Instead of mass cell verticalization, collective twitching initially leads to formation of transient layers of horizontally aligned cells. As the colony expands across the interface, some bacteria are driven upwards, forming new layers above the original monolayer (Fig 5E and supporting video 8). The distribution of pixel intensities in widefield fluorescence images of the colony show 3 distinct peaks, corresponding to the first 3 layers of cells (Fig 5F). Timelapse videos of the system show that layers (including the first one) are dynamic, continuously appearing and disappearing as cells move across the interface (supporting videos 9). This transience is notably different from biofilm formation when bacteria are unconfined, where cells are permanently attached to the surface and 3D structures remain in place. Layer formation first occurs and is most prominent in the protrusions which also drive the expansion of the colony. Unlike under chaotic twitching, cells in these additional layers do not stand up vertically from the surface, but remain aligned parallel to the interface. This particular type of collective behavior has previously been reported, but only in \textit{Myxococcus xanthus} colonies. There, it has been shown that cells are preferentially driven towards $+1/2$ topological defects, causing layer formation. The same mechanism explains the layer formation first shown here in \textit{P. a.} colonies, particularly in groups of highly aligned cells protruding from the edge of the colony and containing a $+1/2$ defect. In \textit{M. xanthus} colonies, it has been shown that layer formation is associated with higher active stresses in the cell monolayer (which tend to arise near $+1/2$ defects)~\cite{Han2025}. The results presented here suggest that such active stresses are responsible for the expansion of the colony via protrusions of highly aligned and collectively moving cells undergoing collective twitching, as shown in Fig 5. When cells move in unison, they push against the glass-TSA interface with greater force, allowing the colony to expand along the interface.

At later times (24 to 48 hours), in colonies undergoing collective twitching, cells far from the edge of the colony become constrained by steric forces from surrounding cells, and thus unable to swarm effectively. At this stage, cells are forced to tilt upwards similarly to those undergoing chaotic twitching, eventually leading to mass verticalization (Fig A2C). Together with the results on chaotic twitching described above (Fig 4), this observation suggests that cell verticalization occurs whenever groups of cells are confined in all three dimensions and unable to expand between the two surfaces. Cells undergoing collective twitching are initially able to avoid this outcome thanks to their coordinated motion, while cells undergoing chaotic twitching and cells far from the colony edge are not. This finding is in partial agreement with previous work, which observed similar mass verticalization (``rosette formation'') under similar conditions only for \textit{pilH} mutants, which swim roughly twice as fast as the WT~\cite{Meacock2020}. By contrast, it is shown here that verticalization of nearly all cells can also occur in the \textit{WT} strain in confined conditions, and that cells in the confined colonies have orientations that are significantly different from those in unconfined (`surficial') biofilms. Even after $48$ hours, cells in unconfined biofilms are randomly oriented relative to the interface, and do not display the dense, highly ordered hexagonal packing found under confinement. Additionally, mass verticalization was observed consistently under relatively stiff ($1.5\%$) agar, which had previously been reported to suppress cell verticalization~\cite{Meacock2020}. 
However, the association between greater forces exerted by the bacteria and their verticalization is consistent with the previous work. Additionally, between glass and TSA, cells undergoing chaotic twitching move faster than those undergoing collective twitching, on average (Fig 3C), suggesting that differences in the swimming speeds of different subpopulations could explain why these two distinct modes of collective behaviors were observed under identical conditions. This explanation is also used to explain why \textit{pilH} mutants became verticalized in the previous work.
The precise cause of the divergence between chaotic and collective twitching requires further investigation. The fact that both were observed under identical conditions (even simultaneously in a single sample), suggests that subtle changes in bacterial behavior, perhaps in response to equally subtle differences in their environment, can have significant effects on colony development and the bacteria's ability to navigate confined spaces.

Overall, these experiments highlight the importance of collective behaviors at the microscale in determining the larger-scale spatiotemporal evolution of bacterial colonies. Differences in geometry (including confinement), material properties, nutrient sources, and gene expression, must all be considered to fully understand how bacteria self-organize to adapt to environmental challenges. Only through high resolution microscopy coupled with quantitative image analysis can the biophysics of bacterial proliferation and the varied morphologies of bacterial colonies be understood. Here, the effect of confinement on the behavior of \textit{P. aeruginosa} is clear. Both in liquid medium and nutrient agar, confinement enhances cell twitching motility, disrupts orientational ordering in cell monolayers, and leads to distinct modes of collective behaviors not observed in surficial colonies. 

\newpage

\section*{Methods}

\subsection*{Bacteria cultures}

\textit{Pseudomonas aeruginosa} (\textit{P. a.}) PAO1 expressing GFP~\cite{Klausen2003} were taken from frozen glycerol stocks and grown overnight in 30\% Tryptich Soy Broth (TSB) containing 0.25\% glucose. The overnight cultures reached an optical density at 600 nm of approximately $1.2$ (late exponential phase) and were subsequently diluted $100\times$ in fresh medium or phosphate-buffered saline (PBS), depending on the experimental condition (see below). 

For unconfined growth on agar, fresh plates of tryptic soy agar (TSA) with an agar concentration of 1.5\% were prepared and autoclaved. Then, 1 \textmu m of the dilute bacterial suspension in PBS was pipetted onto the agar. The liquid was left to evaporate or diffuse into the agar, and the plate was inverted onto a microscope stage for subsequent imaging. 

\subsection*{Microfluidic flow chamber}

A microfluidic flow chamber was constructed as previously described~\cite{Hickl2025} for continuous, direct observation of growing bacterial colonies. In brief, 1 mm thick, rectangular PDMS spacers were placed on a glass coverslip (0.17 mm thick). A 0.1 mm thin film of PDMS was placed on one half of the coverslip. Inlet and outlet tubes were attached to both ends of the flow chamber, and another cover slip was placed on top. The device was sealed with epoxy to prevent leaks. The resulting chamber was initially filled with the dilute suspension of \textit{P. a.} in PBS, resulting in some attachment of cells to both the coverslip and the PDMS film. Then, bacteria were left to grow unconfined under continuous flow of fresh medium at a rate of $0.01$ mL/min. The slow flow rate ensured that adhered bacteria were not removed, but was sufficient to provide nutrients for continued growth, and to progressively remove non-adhered cells. All experiments were conducted at room temperature.

\subsection*{2D confinement of bacteria}

To investigate the growth of bacteria under 2D confinement, $1$ \textmu L of the dilute bacteria suspension in PBS was pipetted onto a glass-bottom petri dish (glass thickness 0.1 mm). The liquid was allowed to evaporate completely. Fresh TSA with an agar concentration of 1.5\% was prepared by microwaving until boiling. The liquid agar was allowed to cool until just above its gelling point (to about 45 $^\circ$C) before 1 mL was pipetted onto the petri dish with the bacteria, causing the TSA to gel almost immediately. This setup allowed bacteria to grow continuously under confinement without the addition of fresh medium. This approach was found to provide much more consistent confinement than simply placing a solid piece of TSA onto the coverslip, which was not sufficient to fully confine the bacteria in 2D. 

To study confinement between glass and PDMS, 5 \textmu L of a dilute bacterial suspension in fresh medium (30\% TSB with 0.25\% glucose) was pipetted onto coverglass. Then, a piece of 1 mm thick PDMS was firmly placed on top of the liquid to confine the cells to a monolayer. Additional liquid was added around the edges of the PDMS to prevent the interface from drying out. During the subsequent imaging, great care was taken to only image cells confined to a single monolayer. Such regions were easy to identify by the absence of a vertical gap between cells attached to the glass and cells attached to the PDMS. 

\subsection*{Imaging}

The resulting dense monolayers of \textit{P. a.} formed in the experiments described above were imaged using a Zeiss LSM780 confocal microscope with a $63\times$ oil immersion objective and a 488 nm laser. The resulting pixel size was 0.085 \textmu m. Videos and timelapses were recorded at various rates (0.0033 to 1 fps) to quantify the behavior of bacteria on timescale from seconds to hours. 

\subsection*{Image segmentation and analysis}

Image segmentation of bacterial monolayers was performed to identify the position, orientation, and shape of each cell within a colony. The segmentation model used here was trained as previously described~\cite{Hickl2025} on synthetic images processed by a cycle generative adversarial network (cycleGAN). This approach allows custom segmentation models to be created quickly for different experimental setups. Both model training and segmentation are performed using Omnipose~\cite{Cutler2022}. For segmentation of bacteria in z-slices of dense 3D colonies (see Fig 4 and Fig S2), a model trained on synthetic images from SyMBac~\cite{Hardo2022} was used, as it outperformed the model trained as described above.

To further improve the results of the segmentation, a post-processing method was developed to automatically detect undersegmented cells (two cells mistakenly identified as a single cell by the segmentation model), and further split the corresponding mask into two cells. In brief, the convexity of each cell in the segmentation mask was measured, and cells below a certain threshold are divided in two along the line that maximizes the convexity of the two resulting cells. This method was primarily used to improve the segmentation quality in images of bacteria grown on thin PDMS films, where the signal-to-noise ratio is lower and the point spread function (PSF) is wider.

Once images are segmented, custom Python programs are used to calculate local packing fractions, nematic order parameters, and microdomain sizes. See the Supporting Information for a detailed description of the methodology used.

\subsection*{Cell tracking}

A custom cell tracking algorithm was developed to track the motion of individual cells in densely-packed monolayers. Since single-cell segmentation was performed on all analyzed images, the full mask of each cell could be used to assist in matching cells across frames in a video, rather than only the cell positions. Cells from consecutive frames were matched using both the overlap between cell masks, and the centroids of the segmented cells. Cell trajectories were then calculated for every cell in a given video, which were analyzed to measure individual cell speeds and mean square displacements (MSDs) under different experimental conditions. See the SI for a detailed description of the algorithm.

\newpage

\begin{appendices}

\section{Extended Data}\label{secA1}

Colonies of \textit{P. a.} were grown in a microfluidic flow chamber and imaged using timelapse confocal microscopy, either directly on a glass cover slip or on a thin layer of PDMS (Fig. 1A). Nearby cell tend align, and the alignment decays exponentially with increasing cell-to-cell distance $r_{ij}$, with a decay constants of $1.72\pm0.04$ \textmu m and $1.91\pm0.04$ for microcolonies grown on PDMS and glass, respectively (Fig. A1C). This decay constant is equivalent to the mean correlation length of cell orientations. Overall, the alignment and its spatial decay appear nearly identical on PDMS and glass.

\begin{figure*}
  \centering
  \includegraphics[width=\linewidth]{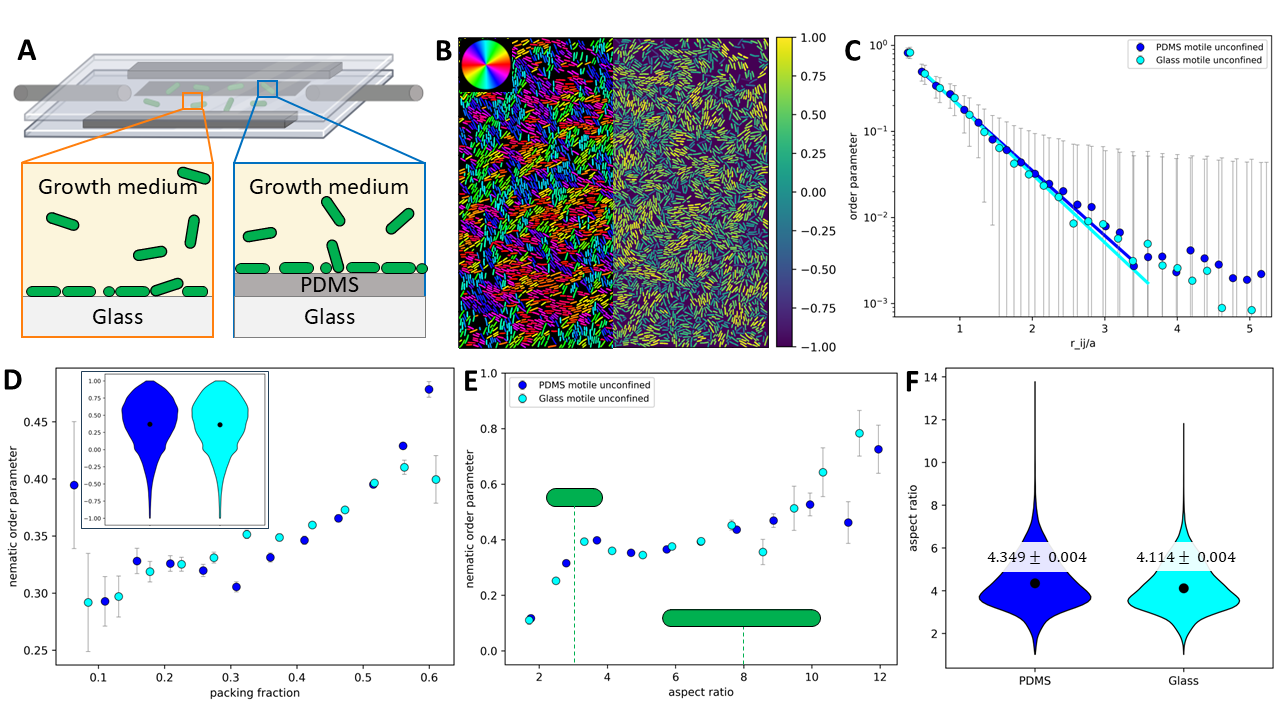}
  \caption{{Unconfined monolayers.} (A) Microfluidic flow chamber for bacterial growth on different surfaces. (B) Representative segmented microcolony, with cells colored by their orientation (left) and the nematic order parameter (right). (C) Cell-to-cell alignment as a function of distance for microcolonies on PDMS and glass. Lines represent exponential decay best fit, and error bars represent the standard error across all cells analyzed in each bin. (D) Single-cell nematic order parameter as a function of cell packing fraction. Inset: Violin plot of nematic order parameters on the two substrates. (E) Nematic order parameter vs. cell aspect ratio. (F) Violin plots of cell aspect ratios on the two substrates. Black dots and values correspond to means.}
  \label{figS1:unconfined}
\end{figure*}

The distribution of $S_i$ is the same for both surfaces (Fig. A1D, inset), with mean values of $0.37$ and $0.36$ on PDMS and glass, respectively. For each cell, the local packing fraction $\phi$-- the proportion of the surface surrounding the cell covered by bacteria -- is calculated. The degree of alignment $S_i$ increases with increasing packing fraction (Fig. A1D). This effect is relatively weak for lower cell densities $\phi< 0.4$, but becomes significantly larger above this threshold. Additionally, the alignment increases with increasing cell aspect ratio $\eta$ (Fig. A1E). 
This effect is strongest for $\eta<4$, nearly disappears for $0.4 < \eta < 8$, and resumes more weakly for $\eta > 8$. 
The distribution of aspect ratios is nearly identical for cells grown on PDMS and glass (Fig A1F). These results show that rod-shaped cells growing unconstrained on various surfaces exhibit significant orientation ordering, which does not depend strongly on the distinct properties of PDMS and glass. In particular, the stiffness of PDMS is much less than that of borosilicate glass (Young's modulus $E\approx1$ MPa compared to $E\approx60$ GPa), and PDMS is much more hydrophobic than glass (mean wetting angles of $100^\circ$ and $27^\circ$, respectively). Both of these parameters have been reported to have a significant effect on bacterial attachment at surfaces, but the self-organization observed here is identical on the two materials. 

\begin{figure*}
  \centering
  \includegraphics[width=\linewidth]{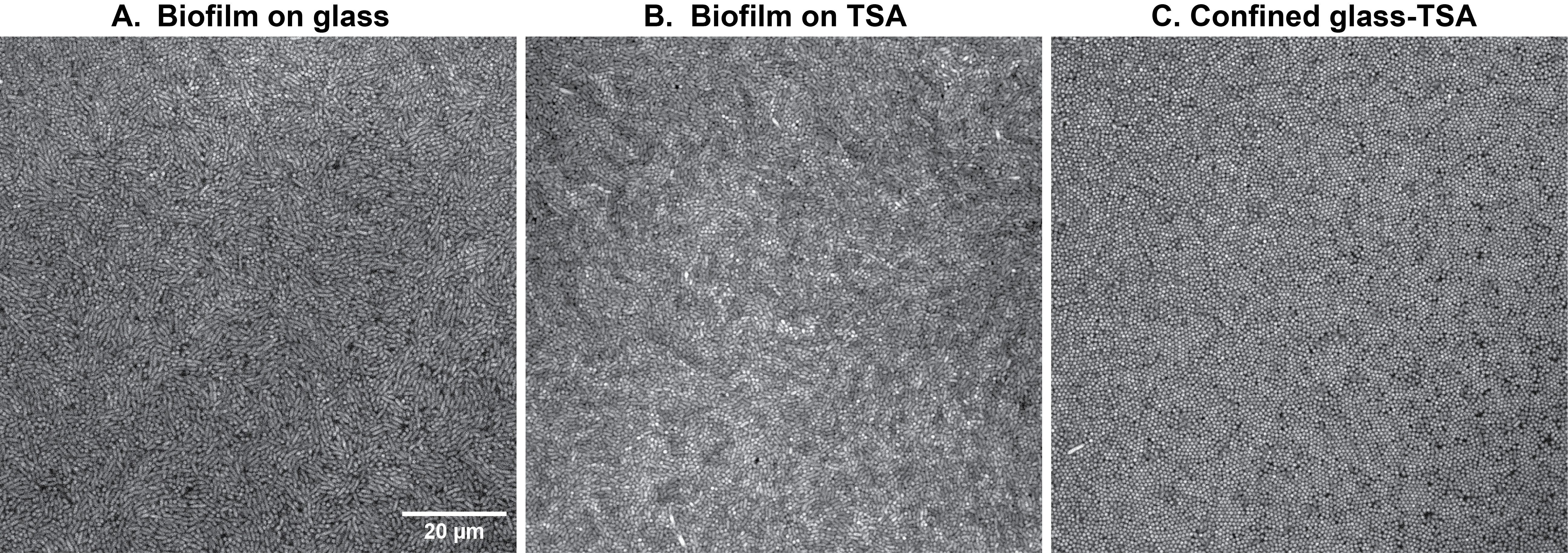}
  \caption{{Biofilm cross-sections.} Slices of unconfined biofilms grown on (A) glass in liquid growth medium TSB or (B) TSA in air, and (C) confined colonies of verticalized cells between glass and TSA. All slices were acquired approximately 2 \textmu m above the surface, and are representative of the images analyzed to create the graph in Fig 4E. The images show the significant effect of confinement on the orientation of cells in dense 3D colonies. Scale bar is 20 \textmu m and applies to all 3 images.}
  \label{figS2:Biofilms}
\end{figure*}

\section{Supporting Videos}
Supporting videos are available at \url{https://doi.org/10.5281/zenodo.17497789}. 

\newpage




\end{appendices}


\bibliography{bibliography}


\begin{thebibliography}{45}
\ifx \bisbn   \undefined \def \bisbn  #1{ISBN #1}\fi
\ifx \binits  \undefined \def \binits#1{#1}\fi
\ifx \bauthor  \undefined \def \bauthor#1{#1}\fi
\ifx \batitle  \undefined \def \batitle#1{#1}\fi
\ifx \bjtitle  \undefined \def \bjtitle#1{#1}\fi
\ifx \bvolume  \undefined \def \bvolume#1{\textbf{#1}}\fi
\ifx \byear  \undefined \def \byear#1{#1}\fi
\ifx \bissue  \undefined \def \bissue#1{#1}\fi
\ifx \bfpage  \undefined \def \bfpage#1{#1}\fi
\ifx \blpage  \undefined \def \blpage #1{#1}\fi
\ifx \burl  \undefined \def \burl#1{\textsf{#1}}\fi
\ifx \doiurl  \undefined \def \doiurl#1{\url{https://doi.org/#1}}\fi
\ifx \betal  \undefined \def \betal{\textit{et al.}}\fi
\ifx \binstitute  \undefined \def \binstitute#1{#1}\fi
\ifx \binstitutionaled  \undefined \def \binstitutionaled#1{#1}\fi
\ifx \bctitle  \undefined \def \bctitle#1{#1}\fi
\ifx \beditor  \undefined \def \beditor#1{#1}\fi
\ifx \bpublisher  \undefined \def \bpublisher#1{#1}\fi
\ifx \bbtitle  \undefined \def \bbtitle#1{#1}\fi
\ifx \bedition  \undefined \def \bedition#1{#1}\fi
\ifx \bseriesno  \undefined \def \bseriesno#1{#1}\fi
\ifx \blocation  \undefined \def \blocation#1{#1}\fi
\ifx \bsertitle  \undefined \def \bsertitle#1{#1}\fi
\ifx \bsnm \undefined \def \bsnm#1{#1}\fi
\ifx \bsuffix \undefined \def \bsuffix#1{#1}\fi
\ifx \bparticle \undefined \def \bparticle#1{#1}\fi
\ifx \barticle \undefined \def \barticle#1{#1}\fi
\bibcommenthead
\ifx \bconfdate \undefined \def \bconfdate #1{#1}\fi
\ifx \botherref \undefined \def \botherref #1{#1}\fi
\ifx \url \undefined \def \url#1{\textsf{#1}}\fi
\ifx \bchapter \undefined \def \bchapter#1{#1}\fi
\ifx \bbook \undefined \def \bbook#1{#1}\fi
\ifx \bcomment \undefined \def \bcomment#1{#1}\fi
\ifx \oauthor \undefined \def \oauthor#1{#1}\fi
\ifx \citeauthoryear \undefined \def \citeauthoryear#1{#1}\fi
\ifx \endbibitem  \undefined \def \endbibitem {}\fi
\ifx \bconflocation  \undefined \def \bconflocation#1{#1}\fi
\ifx \arxivurl  \undefined \def \arxivurl#1{\textsf{#1}}\fi
\csname PreBibitemsHook\endcsname

\bibitem[\protect\citeauthoryear{Dell'Arciprete et~al.}{2018}]{DellArciprete2018}
\begin{barticle}
\bauthor{\bsnm{Dell'Arciprete}, \binits{D.}},
\bauthor{\bsnm{Blow}, \binits{M.L.}},
\bauthor{\bsnm{Brown}, \binits{A.T.}},
\bauthor{\bsnm{Farrell}, \binits{F.D.C.}},
\bauthor{\bsnm{Lintuvuori}, \binits{J.S.}},
\bauthor{\bsnm{McVey}, \binits{A.F.}},
\bauthor{\bsnm{Marenduzzo}, \binits{D.}},
\bauthor{\bsnm{Poon}, \binits{W.C.K.}}:
\batitle{{A growing bacterial colony in two dimensions as an active nematic}}.
\bjtitle{Nature Communications}
\bvolume{9}(\bissue{1}),
\bfpage{1}--\blpage{9}
(\byear{2018})
\doiurl{10.1038/s41467-018-06370-3}
\end{barticle}
\endbibitem

\bibitem[\protect\citeauthoryear{Doostmohammadi et~al.}{2018}]{Doostmohammadi2018}
\begin{barticle}
\bauthor{\bsnm{Doostmohammadi}, \binits{A.}},
\bauthor{\bsnm{Ign{\'{e}}s-Mullol}, \binits{J.}},
\bauthor{\bsnm{Yeomans}, \binits{J.M.}},
\bauthor{\bsnm{Sagu{\'{e}}s}, \binits{F.}}:
\batitle{{Active nematics}}.
\bjtitle{Nature Communications}
\bvolume{9}(\bissue{1}),
\bfpage{3246}
(\byear{2018})
\doiurl{10.1038/s41467-018-05666-8}
\end{barticle}
\endbibitem

\bibitem[\protect\citeauthoryear{Aranson}{2022}]{Aranson2022}
\begin{barticle}
\bauthor{\bsnm{Aranson}, \binits{I.}}:
\batitle{{Bacterial Active Matter}}.
\bjtitle{Reports on Progress in Physics}
(\byear{2022})
\doiurl{10.1088/1361-6633/AC723D}
\end{barticle}
\endbibitem

\bibitem[\protect\citeauthoryear{Costerton et~al.}{1999}]{Costerton1999}
\begin{barticle}
\bauthor{\bsnm{Costerton}, \binits{J.W.}},
\bauthor{\bsnm{Stewart}, \binits{P.S.}},
\bauthor{\bsnm{Greenberg}, \binits{E.P.}}:
\batitle{{Bacterial biofilms: a common cause of persistent infections.}}
\bjtitle{Science}
\bvolume{284}(\bissue{5418}),
\bfpage{1318}--\blpage{22}
(\byear{1999})
\doiurl{10.1126/SCIENCE.284.5418.1318}
\end{barticle}
\endbibitem

\bibitem[\protect\citeauthoryear{Rudrappa et~al.}{2008}]{Rudrappa2008}
\begin{barticle}
\bauthor{\bsnm{Rudrappa}, \binits{T.}},
\bauthor{\bsnm{Biedrzycki}, \binits{M.L.}},
\bauthor{\bsnm{Bais}, \binits{H.P.}}:
\batitle{{Causes and consequences of plant-associated biofilms}}.
\bjtitle{FEMS microbiology ecology}
\bvolume{64}(\bissue{2}),
\bfpage{153}--\blpage{166}
(\byear{2008})
\doiurl{10.1111/J.1574-6941.2008.00465.X}
\end{barticle}
\endbibitem

\bibitem[\protect\citeauthoryear{Flemming}{2020}]{Flemming2020}
\begin{barticle}
\bauthor{\bsnm{Flemming}, \binits{H.C.}}:
\batitle{{Biofouling and me: My Stockholm syndrome with biofilms}}.
\bjtitle{Water Research}
\bvolume{173},
\bfpage{115576}
(\byear{2020})
\doiurl{10.1016/J.WATRES.2020.115576}
\end{barticle}
\endbibitem

\bibitem[\protect\citeauthoryear{Singh et~al.}{2006}]{Singh2006}
\begin{barticle}
\bauthor{\bsnm{Singh}, \binits{R.}},
\bauthor{\bsnm{Paul}, \binits{D.}},
\bauthor{\bsnm{Jain}, \binits{R.K.}}:
\batitle{{Biofilms: implications in bioremediation}}.
\bjtitle{Trends in Microbiology}
\bvolume{14}(\bissue{9}),
\bfpage{389}--\blpage{397}
(\byear{2006})
\doiurl{10.1016/J.TIM.2006.07.001}
\end{barticle}
\endbibitem

\bibitem[\protect\citeauthoryear{Hazen et~al.}{2015}]{Hazen2015}
\begin{barticle}
\bauthor{\bsnm{Hazen}, \binits{T.C.}},
\bauthor{\bsnm{Prince}, \binits{R.C.}},
\bauthor{\bsnm{Mahmoudi}, \binits{N.}}:
\batitle{{Marine Oil Biodegradation}}.
\bjtitle{Environmental Science \& Technology}
\bvolume{50}(\bissue{5}),
\bfpage{2121}--\blpage{2129}
(\byear{2015})
\doiurl{10.1021/acs.est.5b03333}
\end{barticle}
\endbibitem

\bibitem[\protect\citeauthoryear{Hickl et~al.}{2023}]{Hickl2023}
\begin{barticle}
\bauthor{\bsnm{Hickl}, \binits{V.}},
\bauthor{\bsnm{Pamu}, \binits{H.H.}},
\bauthor{\bsnm{Juarez}, \binits{G.}}:
\batitle{{Hydrodynamic Treadmill Reveals Reduced Rising Speeds of Oil Droplets Deformed by Marine Bacteria}}.
\bjtitle{Environmental Science and Technology}
(\byear{2023})
\doiurl{10.1021/ACS.EST.3C04902}
\end{barticle}
\endbibitem

\bibitem[\protect\citeauthoryear{Flemming and Wingender}{2010}]{Flemming2010}
\begin{barticle}
\bauthor{\bsnm{Flemming}, \binits{H.C.}},
\bauthor{\bsnm{Wingender}, \binits{J.}}:
\batitle{{The biofilm matrix}}.
\bjtitle{Nature Reviews Microbiology 2010 8:9}
\bvolume{8}(\bissue{9}),
\bfpage{623}--\blpage{633}
(\byear{2010})
\doiurl{10.1038/nrmicro2415}
\end{barticle}
\endbibitem

\bibitem[\protect\citeauthoryear{Rather et~al.}{2021}]{Rather2021}
\begin{barticle}
\bauthor{\bsnm{Rather}, \binits{M.A.}},
\bauthor{\bsnm{Gupta}, \binits{K.}},
\bauthor{\bsnm{Mandal}, \binits{M.}}:
\batitle{{Microbial biofilm: formation, architecture, antibiotic resistance, and control strategies}}.
\bjtitle{Brazilian Journal of Microbiology}
\bvolume{52}(\bissue{4}),
\bfpage{1701}
(\byear{2021})
\doiurl{10.1007/S42770-021-00624-X}
\end{barticle}
\endbibitem

\bibitem[\protect\citeauthoryear{Sauer et~al.}{2022}]{Sauer2022}
\begin{barticle}
\bauthor{\bsnm{Sauer}, \binits{K.}},
\bauthor{\bsnm{Stoodley}, \binits{P.}},
\bauthor{\bsnm{Goeres}, \binits{D.M.}},
\bauthor{\bsnm{Hall-Stoodley}, \binits{L.}},
\bauthor{\bsnm{Burm{\o}lle}, \binits{M.}},
\bauthor{\bsnm{Stewart}, \binits{P.S.}},
\bauthor{\bsnm{Bjarnsholt}, \binits{T.}}:
\batitle{{The biofilm life cycle: expanding the conceptual model of biofilm formation}}.
\bjtitle{Nature reviews. Microbiology}
\bvolume{20}(\bissue{10}),
\bfpage{608}--\blpage{620}
(\byear{2022})
\doiurl{10.1038/S41579-022-00767-0}
\end{barticle}
\endbibitem

\bibitem[\protect\citeauthoryear{Yan et~al.}{2016}]{Yan2016}
\begin{barticle}
\bauthor{\bsnm{Yan}, \binits{J.}},
\bauthor{\bsnm{Sharo}, \binits{A.G.}},
\bauthor{\bsnm{Stone}, \binits{H.A.}},
\bauthor{\bsnm{Wingreen}, \binits{N.S.}},
\bauthor{\bsnm{Bassler}, \binits{B.L.}}:
\batitle{{Vibrio cholerae biofilm growth program and architecture revealed by single-cell live imaging}}.
\bjtitle{Proceedings of the National Academy of Sciences of the United States of America}
\bvolume{113}(\bissue{36}),
\bfpage{5337}--\blpage{5343}
(\byear{2016})
\doiurl{10.1073/PNAS.1611494113/SUPPL_FILE/PNAS.1611494113.SM08.MOV}
\end{barticle}
\endbibitem

\bibitem[\protect\citeauthoryear{Hartmann et~al.}{2019}]{Hartmann2019}
\begin{barticle}
\bauthor{\bsnm{Hartmann}, \binits{R.}},
\bauthor{\bsnm{Singh}, \binits{P.K.}},
\bauthor{\bsnm{Pearce}, \binits{P.}},
\bauthor{\bsnm{Mok}, \binits{R.}},
\bauthor{\bsnm{Song}, \binits{B.}},
\bauthor{\bsnm{D{\'{i}}az-Pascual}, \binits{F.}},
\bauthor{\bsnm{Dunkel}, \binits{J.}},
\bauthor{\bsnm{Drescher}, \binits{K.}}:
\batitle{{Emergence of three-dimensional order and structure in growing biofilms}}.
\bjtitle{Nature Physics}
\bvolume{15}(\bissue{3}),
\bfpage{251}--\blpage{256}
(\byear{2019})
\doiurl{10.1038/s41567-018-0356-9}
\end{barticle}
\endbibitem

\bibitem[\protect\citeauthoryear{Shimaya and Takeuchi}{2022}]{Shimaya2022}
\begin{barticle}
\bauthor{\bsnm{Shimaya}, \binits{T.}},
\bauthor{\bsnm{Takeuchi}, \binits{K.A.}}:
\batitle{{Tilt-induced polar order and topological defects in growing bacterial populations}}.
\bjtitle{PNAS Nexus}
\bvolume{1}(\bissue{5}),
\bfpage{1}--\blpage{11}
(\byear{2022})
\doiurl{10.1093/PNASNEXUS/PGAC269}
{\href{https://arxiv.org/abs/2106.10954}{{arXiv:2106.10954}}}
\end{barticle}
\endbibitem

\bibitem[\protect\citeauthoryear{Volfson et~al.}{2008}]{Volfson2008}
\begin{barticle}
\bauthor{\bsnm{Volfson}, \binits{D.}},
\bauthor{\bsnm{Cookson}, \binits{S.}},
\bauthor{\bsnm{Hasty}, \binits{J.}},
\bauthor{\bsnm{Tsimring}, \binits{L.S.}}:
\batitle{{Biomechanical ordering of dense cell populations}}.
\bjtitle{Proceedings of the National Academy of Sciences of the United States of America}
\bvolume{105}(\bissue{40}),
\bfpage{15346}--\blpage{15351}
(\byear{2008})
\doiurl{10.1073/PNAS.0706805105/SUPPL_FILE/0706805105SI.PDF}
\end{barticle}
\endbibitem

\bibitem[\protect\citeauthoryear{Beroz et~al.}{2018}]{Beroz2018}
\begin{barticle}
\bauthor{\bsnm{Beroz}, \binits{F.}},
\bauthor{\bsnm{Yan}, \binits{J.}},
\bauthor{\bsnm{Meir}, \binits{Y.}},
\bauthor{\bsnm{Sabass}, \binits{B.}},
\bauthor{\bsnm{Stone}, \binits{H.A.}},
\bauthor{\bsnm{Bassler}, \binits{B.L.}},
\bauthor{\bsnm{Wingreen}, \binits{N.S.}}:
\batitle{{Verticalization of bacterial biofilms}}.
\bjtitle{Nature Physics}
\bvolume{14}(\bissue{9}),
\bfpage{954}--\blpage{960}
(\byear{2018})
\doiurl{10.1038/s41567-018-0170-4}
\end{barticle}
\endbibitem

\bibitem[\protect\citeauthoryear{Meacock et~al.}{2020}]{Meacock2020}
\begin{barticle}
\bauthor{\bsnm{Meacock}, \binits{O.J.}},
\bauthor{\bsnm{Doostmohammadi}, \binits{A.}},
\bauthor{\bsnm{Foster}, \binits{K.R.}},
\bauthor{\bsnm{Yeomans}, \binits{J.M.}},
\bauthor{\bsnm{Durham}, \binits{W.M.}}:
\batitle{{Bacteria solve the problem of crowding by moving slowly}}.
\bjtitle{Nature Physics 2020 17:2}
\bvolume{17}(\bissue{2}),
\bfpage{205}--\blpage{210}
(\byear{2020})
\doiurl{10.1038/s41567-020-01070-6}
{\href{https://arxiv.org/abs/2008.07915}{{arXiv:2008.07915}}}
\end{barticle}
\endbibitem

\bibitem[\protect\citeauthoryear{Copenhagen et~al.}{2020}]{Copenhagen2020}
\begin{barticle}
\bauthor{\bsnm{Copenhagen}, \binits{K.}},
\bauthor{\bsnm{Alert}, \binits{R.}},
\bauthor{\bsnm{Wingreen}, \binits{N.S.}},
\bauthor{\bsnm{Shaevitz}, \binits{J.W.}}:
\batitle{{Topological defects promote layer formation in Myxococcus xanthus colonies}}.
\bjtitle{Nature Physics}
\bvolume{17}(\bissue{2}),
\bfpage{211}--\blpage{215}
(\byear{2020})
\doiurl{10.1038/s41567-020-01056-4}
{\href{https://arxiv.org/abs/2001.03804}{{arXiv:2001.03804}}}
\end{barticle}
\endbibitem

\bibitem[\protect\citeauthoryear{Han et~al.}{2025}]{Han2025}
\begin{barticle}
\bauthor{\bsnm{Han}, \binits{E.}},
\bauthor{\bsnm{Fei}, \binits{C.}},
\bauthor{\bsnm{Alert}, \binits{R.}},
\bauthor{\bsnm{Copenhagen}, \binits{K.}},
\bauthor{\bsnm{Koch}, \binits{M.D.}},
\bauthor{\bsnm{Wingreen}, \binits{N.S.}},
\bauthor{\bsnm{Shaevitz}, \binits{J.W.}}:
\batitle{{Local polar order controls mechanical stress and triggers layer formation in Myxococcus xanthus colonies}}.
\bjtitle{Nature Communications 2025 16:1}
\bvolume{16}(\bissue{1}),
\bfpage{1}--\blpage{10}
(\byear{2025})
\doiurl{10.1038/s41467-024-55806-6}
\end{barticle}
\endbibitem

\bibitem[\protect\citeauthoryear{Liu et~al.}{2019}]{Liu2019}
\begin{barticle}
\bauthor{\bsnm{Liu}, \binits{G.}},
\bauthor{\bsnm{Patch}, \binits{A.}},
\bauthor{\bsnm{Bahar}, \binits{F.}},
\bauthor{\bsnm{Yllanes}, \binits{D.}},
\bauthor{\bsnm{Welch}, \binits{R.D.}},
\bauthor{\bsnm{Marchetti}, \binits{M.C.}},
\bauthor{\bsnm{Thutupalli}, \binits{S.}},
\bauthor{\bsnm{Shaevitz}, \binits{J.W.}}:
\batitle{{Self-Driven Phase Transitions Drive Myxococcus xanthus Fruiting Body Formation}}.
\bjtitle{Physical Review Letters}
\bvolume{122}(\bissue{24}),
\bfpage{248102}
(\byear{2019})
\doiurl{10.1103/PHYSREVLETT.122.248102/MOVIES4.MOV}
{\href{https://arxiv.org/abs/1709.06012}{{arXiv:1709.06012}}}
\end{barticle}
\endbibitem

\bibitem[\protect\citeauthoryear{Grobas et~al.}{2021}]{Grobas2021}
\begin{botherref}
\oauthor{\bsnm{Grobas}, \binits{I.}},
\oauthor{\bsnm{Polin}, \binits{M.}},
\oauthor{\bsnm{Asally}, \binits{M.}}:
{Swarming bacteria undergo localized dynamic phase transition to form stress-induced biofilms}.
eLife
\textbf{10}
(2021)
\doiurl{10.7554/ELIFE.62632}
\end{botherref}
\endbibitem

\bibitem[\protect\citeauthoryear{You et~al.}{2018}]{You2018}
\begin{barticle}
\bauthor{\bsnm{You}, \binits{Z.}},
\bauthor{\bsnm{Pearce}, \binits{D.J.G.}},
\bauthor{\bsnm{Sengupta}, \binits{A.}},
\bauthor{\bsnm{Giomi}, \binits{L.}}:
\batitle{{Geometry and Mechanics of Microdomains in Growing Bacterial Colonies}}.
\bjtitle{Physical Review X}
\bvolume{8}(\bissue{3}),
\bfpage{031065}
(\byear{2018})
\doiurl{10.1103/PHYSREVX.8.031065/FIGURES/8/MEDIUM}
{\href{https://arxiv.org/abs/1703.04504}{{arXiv:1703.04504}}}
\end{barticle}
\endbibitem

\bibitem[\protect\citeauthoryear{Nagel et~al.}{2020}]{Nagel2020}
\begin{barticle}
\bauthor{\bsnm{Nagel}, \binits{A.M.}},
\bauthor{\bsnm{Greenberg}, \binits{M.}},
\bauthor{\bsnm{Shendruk}, \binits{T.N.}},
\bauthor{\bsnm{Haan}, \binits{H.W.}}:
\batitle{{Collective Dynamics of Model Pili-Based Twitcher-Mode Bacilliforms}}.
\bjtitle{Scientific Reports 2020 10:1}
\bvolume{10}(\bissue{1}),
\bfpage{1}--\blpage{16}
(\byear{2020})
\doiurl{10.1038/s41598-020-67212-1}
\end{barticle}
\endbibitem

\bibitem[\protect\citeauthoryear{You et~al.}{2021}]{You2021}
\begin{botherref}
\oauthor{\bsnm{You}, \binits{Z.}},
\oauthor{\bsnm{Pearce}, \binits{D.J.G.}},
\oauthor{\bsnm{Giomi}, \binits{L.}}:
{Confinement-induced self-organization in growing bacterial colonies}.
Science Advances
\textbf{7}(4)
(2021)
\doiurl{10.1126/SCIADV.ABC8685/SUPPL_FILE/ABC8685_SM.PDF}
{\href{https://arxiv.org/abs/2004.14890}{{arXiv:2004.14890}}}
\end{botherref}
\endbibitem

\bibitem[\protect\citeauthoryear{Langeslay and Juarez}{2023}]{Langeslay2023a}
\begin{barticle}
\bauthor{\bsnm{Langeslay}, \binits{B.}},
\bauthor{\bsnm{Juarez}, \binits{G.}}:
\batitle{{Microdomains and stress distributions in bacterial monolayers on curved interfaces}}.
\bjtitle{Soft Matter}
\bvolume{19}(\bissue{20}),
\bfpage{3605}--\blpage{3613}
(\byear{2023})
\doiurl{10.1039/D2SM01498J}
{\href{https://arxiv.org/abs/2212.00233}{{arXiv:2212.00233}}}
\end{barticle}
\endbibitem

\bibitem[\protect\citeauthoryear{Bechinger et~al.}{2016}]{Bechinger2016}
\begin{barticle}
\bauthor{\bsnm{Bechinger}, \binits{C.}},
\bauthor{\bsnm{{Di Leonardo}}, \binits{R.}},
\bauthor{\bsnm{L{\"{o}}wen}, \binits{H.}},
\bauthor{\bsnm{Reichhardt}, \binits{C.}},
\bauthor{\bsnm{Volpe}, \binits{G.}},
\bauthor{\bsnm{Volpe}, \binits{G.}}:
\batitle{{Active particles in complex and crowded environments}}.
\bjtitle{Reviews of Modern Physics}
\bvolume{88}(\bissue{4}),
\bfpage{045006}
(\byear{2016})
\doiurl{10.1103/RevModPhys.88.045006}
{\href{https://arxiv.org/abs/1602.00081}{{arXiv:1602.00081}}}
\end{barticle}
\endbibitem

\bibitem[\protect\citeauthoryear{Arnaouteli et~al.}{2021}]{Arnaouteli2021}
\begin{barticle}
\bauthor{\bsnm{Arnaouteli}, \binits{S.}},
\bauthor{\bsnm{Bamford}, \binits{N.C.}},
\bauthor{\bsnm{Stanley-Wall}, \binits{N.R.}},
\bauthor{\bsnm{Kov{\'{a}}cs}, \binits{{\'{A}}.T.}}:
\batitle{{Bacillus subtilis biofilm formation and social interactions}}.
\bjtitle{Nature Reviews Microbiology 2021 19:9}
\bvolume{19}(\bissue{9}),
\bfpage{600}--\blpage{614}
(\byear{2021})
\doiurl{10.1038/s41579-021-00540-9}
\end{barticle}
\endbibitem

\bibitem[\protect\citeauthoryear{Moore-Ott et~al.}{2022}]{Moore-Ott2022}
\begin{botherref}
\oauthor{\bsnm{Moore-Ott}, \binits{J.A.}},
\oauthor{\bsnm{Chiu}, \binits{S.}},
\oauthor{\bsnm{Amchin}, \binits{D.B.}},
\oauthor{\bsnm{Bhattacharjee}, \binits{T.}},
\oauthor{\bsnm{Datta}, \binits{S.S.}}:
{A biophysical threshold for biofilm formation}.
eLife
\textbf{11}
(2022)
\doiurl{10.7554/ELIFE.76380}
{\href{https://arxiv.org/abs/2112.02683}{{arXiv:2112.02683}}}
\end{botherref}
\endbibitem

\bibitem[\protect\citeauthoryear{Cho et~al.}{2007}]{Cho2007}
\begin{barticle}
\bauthor{\bsnm{Cho}, \binits{H.J.}},
\bauthor{\bsnm{J{\"{o}}nsson}, \binits{H.}},
\bauthor{\bsnm{Campbell}, \binits{K.}},
\bauthor{\bsnm{Melke}, \binits{P.}},
\bauthor{\bsnm{Williams}, \binits{J.W.}},
\bauthor{\bsnm{Jedynak}, \binits{B.}},
\bauthor{\bsnm{Stevens}, \binits{A.M.}},
\bauthor{\bsnm{Groisman}, \binits{A.}},
\bauthor{\bsnm{Levchenko}, \binits{A.}}:
\batitle{{Self-Organization in High-Density Bacterial Colonies: Efficient Crowd Control}}.
\bjtitle{PLOS Biology}
\bvolume{5}(\bissue{11}),
\bfpage{302}
(\byear{2007})
\doiurl{10.1371/JOURNAL.PBIO.0050302}
\end{barticle}
\endbibitem

\bibitem[\protect\citeauthoryear{Bhattacharjee and Datta}{2019}]{Bhattacharjee2019}
\begin{barticle}
\bauthor{\bsnm{Bhattacharjee}, \binits{T.}},
\bauthor{\bsnm{Datta}, \binits{S.S.}}:
\batitle{{Bacterial hopping and trapping in porous media}}.
\bjtitle{Nature Communications 2019 10:1}
\bvolume{10}(\bissue{1}),
\bfpage{1}--\blpage{9}
(\byear{2019})
\doiurl{10.1038/s41467-019-10115-1}
\end{barticle}
\endbibitem

\bibitem[\protect\citeauthoryear{Chen et~al.}{2024}]{Chen2024}
\begin{botherref}
\oauthor{\bsnm{Chen}, \binits{X.}},
\oauthor{\bsnm{Zhang}, \binits{R.}},
\oauthor{\bsnm{Yuan}, \binits{J.}}:
{Vertical confinement enhances surface exploration in bacterial twitching motility}.
Environmental Microbiology
\textbf{26}(7)
(2024)
\doiurl{10.1111/1462-2920.16679}
\end{botherref}
\endbibitem

\bibitem[\protect\citeauthoryear{O'Toole and Kolter}{1998}]{OToole1998}
\begin{barticle}
\bauthor{\bsnm{O'Toole}, \binits{G.A.}},
\bauthor{\bsnm{Kolter}, \binits{R.}}:
\batitle{{Flagellar and twitching motility are necessary for Pseudomonas aeruginosa biofilm development}}.
\bjtitle{Molecular Microbiology}
\bvolume{30}(\bissue{2}),
\bfpage{295}--\blpage{304}
(\byear{1998})
\doiurl{10.1046/J.1365-2958.1998.01062.X}
\end{barticle}
\endbibitem

\bibitem[\protect\citeauthoryear{Weitz et~al.}{2015}]{Weitz2015}
\begin{barticle}
\bauthor{\bsnm{Weitz}, \binits{S.}},
\bauthor{\bsnm{Deutsch}, \binits{A.}},
\bauthor{\bsnm{Peruani}, \binits{F.}}:
\batitle{{Self-propelled rods exhibit a phase-separated state characterized by the presence of active stresses and the ejection of polar clusters}}.
\bjtitle{Physical Review E - Statistical, Nonlinear, and Soft Matter Physics}
\bvolume{92}(\bissue{1}),
\bfpage{012322}
(\byear{2015})
\doiurl{10.1103/PHYSREVE.92.012322/ETA0_3-KAPPA4-N80000.MP4}
{\href{https://arxiv.org/abs/1507.00948}{{arXiv:1507.00948}}}
\end{barticle}
\endbibitem

\bibitem[\protect\citeauthoryear{Sheats et~al.}{2017}]{Sheats2017}
\begin{botherref}
\oauthor{\bsnm{Sheats}, \binits{J.}},
\oauthor{\bsnm{Sclavi}, \binits{B.}},
\oauthor{\bsnm{Lagomarsino}, \binits{M.C.}},
\oauthor{\bsnm{Cicuta}, \binits{P.}},
\oauthor{\bsnm{Dorfman}, \binits{K.D.}}:
{Role of growth rate on the orientational alignment of Escherichia coli in a slit}.
Royal Society Open Science
\textbf{4}(6)
(2017)
\doiurl{10.1098/RSOS.170463}
\end{botherref}
\endbibitem

\bibitem[\protect\citeauthoryear{Onsager}{1949}]{Onsager1949}
\begin{barticle}
\bauthor{\bsnm{Onsager}, \binits{L.}}:
\batitle{{THE EFFECTS OF SHAPE ON THE INTERACTION OF COLLOIDAL PARTICLES}}.
\bjtitle{Annals of the New York Academy of Sciences}
\bvolume{51}(\bissue{4}),
\bfpage{627}--\blpage{659}
(\byear{1949})
\doiurl{10.1111/J.1749-6632.1949.TB27296.X}
\end{barticle}
\endbibitem

\bibitem[\protect\citeauthoryear{Flory}{1956}]{Flory1956}
\begin{barticle}
\bauthor{\bsnm{Flory}, \binits{P.J.}}:
\batitle{{Phase equilibria in solutions of rod-like particles}}.
\bjtitle{Proceedings of the Royal Society of London. Series A. Mathematical and Physical Sciences}
\bvolume{234}(\bissue{1196}),
\bfpage{73}--\blpage{89}
(\byear{1956})
\doiurl{10.1098/RSPA.1956.0016}
\end{barticle}
\endbibitem

\bibitem[\protect\citeauthoryear{Henrici}{1928}]{Henrici1928}
\begin{bbook}
\bauthor{\bsnm{Henrici}, \binits{A.T.}}:
\bbtitle{{Morphologic Variation and the Rate of Growth of Bacteria}}.
\bpublisher{C. C. Thomas},
\blocation{Springfield, IL and Baltimore, MD}
(\byear{1928})
\end{bbook}
\endbibitem

\bibitem[\protect\citeauthoryear{Steinberger et~al.}{2002}]{Steinberger2002}
\begin{barticle}
\bauthor{\bsnm{Steinberger}, \binits{R.E.}},
\bauthor{\bsnm{Allen}, \binits{A.R.}},
\bauthor{\bsnm{Hansma}, \binits{H.G.}},
\bauthor{\bsnm{Holden}, \binits{P.A.}}:
\batitle{{Elongation correlates with nutrient deprivation in Pseudomonas aeruginosa - Unsaturated biofilms}}.
\bjtitle{Microbial Ecology}
\bvolume{43}(\bissue{4}),
\bfpage{416}--\blpage{423}
(\byear{2002})
\doiurl{10.1007/S00248-001-1063-Z}
\end{barticle}
\endbibitem

\bibitem[\protect\citeauthoryear{Ojkic and Banerjee}{2021}]{Ojkic2021}
\begin{barticle}
\bauthor{\bsnm{Ojkic}, \binits{N.}},
\bauthor{\bsnm{Banerjee}, \binits{S.}}:
\batitle{{Bacterial cell shape control by nutrient-dependent synthesis of cell division inhibitors}}.
\bjtitle{Biophysical Journal}
\bvolume{120}(\bissue{11}),
\bfpage{2079}--\blpage{2084}
(\byear{2021})
\doiurl{10.1016/J.BPJ.2021.04.001}
\end{barticle}
\endbibitem

\bibitem[\protect\citeauthoryear{Kayser and Ravech{\'{e}}}{1978}]{Kayser1978}
\begin{barticle}
\bauthor{\bsnm{Kayser}, \binits{R.F.}},
\bauthor{\bsnm{Ravech{\'{e}}}, \binits{H.J.}}:
\batitle{{Bifurcation in Onsager's model of the isotropic-nematic transition}}.
\bjtitle{Physical Review A}
\bvolume{17}(\bissue{6}),
\bfpage{2067}
(\byear{1978})
\doiurl{10.1103/PhysRevA.17.2067}
\end{barticle}
\endbibitem

\bibitem[\protect\citeauthoryear{Klausen et~al.}{2003}]{Klausen2003}
\begin{barticle}
\bauthor{\bsnm{Klausen}, \binits{M.}},
\bauthor{\bsnm{Heydorn}, \binits{A.}},
\bauthor{\bsnm{Ragas}, \binits{P.}},
\bauthor{\bsnm{Lambertsen}, \binits{L.}},
\bauthor{\bsnm{Aaes-J{\o}rgensen}, \binits{A.}},
\bauthor{\bsnm{Molin}, \binits{S.}},
\bauthor{\bsnm{Tolker-Nielsen}, \binits{T.}}:
\batitle{{Biofilm formation by Pseudomonas aeruginosa wild type, flagella and type IV pili mutants}}.
\bjtitle{Molecular Microbiology}
\bvolume{48}(\bissue{6}),
\bfpage{1511}--\blpage{1524}
(\byear{2003})
\doiurl{10.1046/J.1365-2958.2003.03525.X}
\end{barticle}
\endbibitem

\bibitem[\protect\citeauthoryear{Hickl et~al.}{2025}]{Hickl2025}
\begin{barticle}
\bauthor{\bsnm{Hickl}, \binits{V.}},
\bauthor{\bsnm{Khan}, \binits{A.}},
\bauthor{\bsnm{Rossi}, \binits{R.M.}},
\bauthor{\bsnm{{B Silva}}, \binits{B.F.}},
\bauthor{\bsnm{Maniura-Weber}, \binits{K.}}:
\batitle{{Segmentation of dense and multi-species bacterial colonies using models trained on synthetic microscopy images}}.
\bjtitle{PLOS Computational Biology}
\bvolume{21}(\bissue{4}),
\bfpage{1012874}
(\byear{2025})
\doiurl{10.1371/JOURNAL.PCBI.1012874}
{\href{https://arxiv.org/abs/2405.12407}{{arXiv:2405.12407}}}
\end{barticle}
\endbibitem

\bibitem[\protect\citeauthoryear{Cutler et~al.}{2022}]{Cutler2022}
\begin{barticle}
\bauthor{\bsnm{Cutler}, \binits{K.J.}},
\bauthor{\bsnm{Stringer}, \binits{C.}},
\bauthor{\bsnm{Lo}, \binits{T.W.}},
\bauthor{\bsnm{Rappez}, \binits{L.}},
\bauthor{\bsnm{Stroustrup}, \binits{N.}},
\bauthor{\bsnm{{Brook Peterson}}, \binits{S.}},
\bauthor{\bsnm{Wiggins}, \binits{P.A.}},
\bauthor{\bsnm{Mougous}, \binits{J.D.}}:
\batitle{{Omnipose: a high-precision morphology-independent solution for bacterial cell segmentation}}.
\bjtitle{Nature Methods}
\bvolume{19}(\bissue{11}),
\bfpage{1438}--\blpage{1448}
(\byear{2022})
\doiurl{10.1038/s41592-022-01639-4}
\end{barticle}
\endbibitem

\bibitem[\protect\citeauthoryear{Hardo et~al.}{2022}]{Hardo2022}
\begin{barticle}
\bauthor{\bsnm{Hardo}, \binits{G.}},
\bauthor{\bsnm{Noka}, \binits{M.}},
\bauthor{\bsnm{Bakshi}, \binits{S.}}:
\batitle{{Synthetic Micrographs of Bacteria (SyMBac) allows accurate segmentation of bacterial cells using deep neural networks}}.
\bjtitle{BMC Biology}
\bvolume{20}(\bissue{1}),
\bfpage{1}--\blpage{16}
(\byear{2022})
\doiurl{10.1186/S12915-022-01453-6/TABLES/3}
\end{barticle}
\endbibitem

\end{thebibliography}

\end{document}